\documentstyle[11pt,newpasp,twoside,epsf]{article}
\def\edcomment#1{\iffalse\marginpar{\raggedright\sl#1\/}\else\relax\fi}
\reversemarginpar

\begin{document}
\title{Can quantum theory explain dark matter?}
\author{A. D. Ernest}
\affil{University of New England, Armidale, NSW, Australia, 2350}

\begin{abstract}
Certain solutions to a gravitational form of Schrodinger's
equation can yield stable, macroscopic eigenstate solutions having
no classical analogue, with properties resembling those of dark
matter. Some more tractable solutions show: (1) radiative
lifetimes far exceeding the universe's age, implying negligible
emission and inherent stability w.r.t. gravitational collapse, (2)
negligible interaction with EMR and visible matter, (3) potential
to give rise to flat rotation curves and (4) eigenstate energies
and `sizes' consistent with that expected for the galactic halo.
Traditional baryonic particles occupying such eigenstates will be
invisible and weakly interacting, and may be assimilated into
galactic evolution scenarios without significantly disturbing BBN
ratios. It is proposed that such structures may explain the nature
and origin of dark matter.
\end{abstract}

\section{Introduction}
Nesvizhevsky et al.(2002) have experimentally demonstrated the
quantized nature of gravity and thus applying Schr\"{o}dinger's
equation to gravity should be valid in regions \(r>>\)
Schwartzschild radius \(r_s\).  There is nothing in quantum theory
that forbids the existence of a `macroscopic eigenstructure'
formed from a plethora of the gravitational eigenstate solutions,
populated with traditional particles, around a large central
potential such as a massive primordial black hole (MPBH) \(\geq
10^{35}\) kg, as predicted by Ashfordi, N., McDonald, P. \&
Spergel, D. N. (2003), the structure size limited only by the
energy of the highest quantum eigenstate \(E_n\) approaching a
suitably defined minimum binding energy. It is proposed that such
structures might explain the nature and origin of dark matter, and
form the `wimp-like' skeletal basis of galaxies and clusters. Note
that many experiments demonstrate a variety of macroscopic quantum
effects (see for example Friedman et al. 2000) and superluminally
connected quantum systems, macroscopically entangled over many
kilometers (Zbinden et al. 2000).

\section{Eigenstate Radiative Lifetime, Stability, Particle and Photon Interactions, Energy and Size, Density Profile}

Relatively pure, high \(n,\ell\) eigenstate solutions (large \(n\)
Laguerre polynomials with (\(\ell_{min}(\equiv n-p_{max})\leq
\ell\leq n-1\) and \(n\leq 10^{34}\)) have all the properties
required for dark matter (Ernest 2001). \(p_{max} \sim 10 \) gives
a sufficient number of eigenstates to accommodate the mass of dark
matter in a galactic halo. Analysis shows:

(1) Radiative lifetime \(\tau\) (\(= (3\varepsilon_0 \pi \hbar
c^3)/(\omega^3 p_{if}^2)\) where \(p_{if}\equiv\) dipole matrix
element \(\langle i |e \textrm{\textbf{r}} |f\rangle\)) can be \(>
10^6 t_0\) (\(t_0\equiv\) age of the universe) when \(\ell\sim
n\), resulting in long term structural stability, and inability to
further gravitationally collapse.

(2) Low interaction rates with both particles and photons, which
arises from the overlap integrals being negligible unless the
state energy differences are virtually zero. This occurs because
high \(n, \ell\) states are closely spaced and, unlike their
atomic counterparts, are sharply truncated and exhibit highly
symmetric oscillations capable of causing effective cancellation
whenever \((\ell, n)_{final}\) differs sufficiently from \((\ell,
n)_{initial}\). For example, bound inelastic Compton scattering,
the relevant inelastic scattering process here, is negligible
through the overlap integral \(\langle f|
\textbf{\textrm{e}}^{\imath(\textrm{\textbf{k}}_i -
\textrm{\textbf{k}}_f )\cdot \textrm{\textbf{r}}} |i\rangle\)
(Jung 2000), unless \(\textrm{\textbf{k}}_i \approx
\textrm{\textbf{k}}_f \). Similar overlap integral arguments apply
for stimulated transitions and most interactions between
traditional `visible' particles and those in relatively pure high
\(n, \ell\) eigenstates. Note also that the \(p_{max} < 10 \)
eigenstate fraction is negligible in `localized' particles.

(3) If high \(n\) states with \(\ell_{min}\leq \ell\leq n-1\) and
their z-projection sublevels are all filled, \(1/r^2\) radial
density profiles result, leading to flat rotation curves at outer
galactic radii. Formation scenarios incorporating universal
expansion and particle dynamics also suggest the possibility of
shallower profiles at low r.

(4) A central potential mass \(\sim 10^{42}\) kg gives an
eigenstate radius \(r_n\) (\(\approx b_0 n^2\), \(b_0 \equiv\)
Bohr radius equivalent) \(\sim10^{22}\) m, coincident with \(E_n\)
approaching a suitably defined approximate minimum binding energy.

\section{Discussion and Formation Scenario}

Eigenstructures formed around MPBHs through a process of
`gravitational recombination', the fraction of matter ending up in
relatively pure eigenstates depending on rate of universal cooling
versus rate of expansion, and the rate at which matter could be
retained in the concurrently forming gravitational wells. It
appears that the MPBH mass is critical in determining the filling
rate and hence final mass achievable before the recombination rate
peters out due to expansion. Rough estimates suggest
eigenstructure masses of \(> 10^{42}\) kg are achievable with MPBH
masses of \(7 \times 10^{35}\) kg. Galaxies formed later via
standard gravitational collapse of the residual matter. Rough
estimates suggest recombination ceased with 90 s and BBN ratios
would not have been effected. In any case, the expected deviations
from accepted BBN ratios in the high density asymmetry regions
caused by MPBH potentials and formation process would be `buried'
in the eigenstructure and `invisible' to modern BBN ratio
measurements.

\end{document}